%% file: main_elsevier.tex
\documentclass[final,3p,times]{elsarticle}

\usepackage{amsmath,amsfonts,amssymb}
\usepackage{mathtools}
\usepackage{algorithmic}
\usepackage{xcolor}
\usepackage{tikz}
\usepackage{booktabs}
\usepackage{algorithm}
\usepackage{changes}
\usepackage{array}
\usepackage[caption=false,font=normalsize,labelfont=sf,textfont=sf]{subfig}
\usepackage{textcomp}
\usepackage{stfloats}
\usepackage{url}
\usepackage{verbatim}
\usepackage{graphicx}
\usepackage{cite}

\definecolor{bluecolorblind}{rgb}{0.0039, 0.451, 0.698}
\definecolor{tabblue}{rgb}{0.121, 0.466, 0.705}
\usepackage[colorlinks=true,linkcolor=tabblue,urlcolor=tabblue,citecolor=tabblue]{hyperref}

\def\ie{\textit{i.e.}}
\def\eg{\textit{e.g.}}

\journal{Journal of Computational Science}

\begin{document}

\begin{frontmatter}



\title{Efficient Portfolio Selection through Preference Aggregation
  with Quicksort and the Bradley--Terry Model}


\author{Yurun Ge$^{a,b}$, Lucas B\"ottcher$^{c,d}$, Tom Chou$^{b,e}$, Maria R. D'Orsogna$^{a,b}$\corref{cor1}}


\affiliation{organization={Department of Mathematics, California State University at Northridge},
            city={Los Angeles},
            state={CA},
            country={USA}}

\affiliation{organization={Department of Computational Medicine, University of California, Los Angeles},
            city={Los Angeles},
            state={CA},
            country={USA}}

\affiliation{organization={Department of Computational Science and Philosophy, Frankfurt School of Finance and Management},
            city={Frankfurt a. M.},
            country={Germany}}

\affiliation{organization={Laboratory for Systems Medicine, University of Florida},
            city={Gainesville},
            state={FL},
            country={USA}}

\affiliation{organization={Department of Mathematics, University of California, Los Angeles},
            city={Los Angeles},
            state={CA},
            country={USA}}

\cortext[cor1]{dorsogna@csun.edu}
\begin{abstract}
How to allocate limited resources to projects that will yield the
greatest long-term benefits is a problem that often arises in
decision-making under uncertainty. For example, organizations may need
to evaluate and select innovation projects with risky returns.
Similarly, when allocating resources to research projects, funding
agencies are tasked with identifying the most promising proposals
based on idiosyncratic criteria. Finally, in participatory budgeting,
a local community may need to select a subset of public projects to
fund. Regardless of context, agents must estimate the uncertain values
of a potentially large number of projects. Developing parsimonious
methods to compare these projects, and aggregating agent evaluations
so that the overall benefit is maximized, are critical in assembling
the best project portfolio. Unlike in standard sorting algorithms,
evaluating projects on the basis of uncertain long-term benefits
introduces additional complexities. We propose comparison rules based
on Quicksort and the Bradley--Terry model, which connects rankings to
pairwise ``win'' probabilities.  In our model, each agent determines
win probabilities of a pair of projects based on his or her specific
evaluation of the projects' long-term benefit. The win probabilities
are then appropriately aggregated and used to rank projects. Several
of the methods we propose perform better than the two most effective
aggregation methods currently available.  Additionally, our methods
can be combined with sampling techniques to significantly reduce the
number of pairwise comparisons. We also discuss how the Bradley--Terry
portfolio selection approach can be implemented in practice.
\end{abstract}




\end{frontmatter}



\section{Introduction}
The problem of allocating limited resources to select
projects that offer the greatest benefit to stakeholders arises in
many decision-making tasks.  One of the main issues in project
selection under uncertainty that it is often difficult to estimate the
long-term benefit (or ``value") of any given project, resulting in a
large heterogeneity of estimates when multiple agents are queried.
One example is project selection in organizational contexts, where
multiple members are called to evaluate a number of innovation
projects with uncertain
returns~\citep{csaszar2013organizational,boettcher2024selection}. The
goal is to select the most promising options while balancing the
diverse member inputs.  A related example is participatory
budgeting~\citep{Wampler2007,Aziz2021,benade2021preference,helbing2023democracy,yang2024designing}, where
communities must choose which public projects should receive
funding. Regardless of the specific context, agents must evaluate the
benefit associated with each of a large set of projects under
conditions of uncertainty. Given the uncertain and heterogeneous
inputs from various stakeholders, devising effective comparison and
aggregation methods is crucial for reducing cognitive load on agents
while selecting an optimal portfolio.

Prior research has examined the effectiveness of various aggregation
methods, such as voting, averaging, and delegation to experts. An
existing model of organizational
decision-making~\citep{csaszar2013organizational} assumes that agents
evaluate and approve one project at a time, without
comparisons. Projects are characterized by both type and intrinsic
value, while agents' evaluations depend on their specific expertise
with regard to the project types. This model has been extended in
\citep{boettcher2024selection} to address portfolio selection under
budget constraints. Evaluating multiple projects with uncertain values
and choosing a subset considered most valuable by agents is closely
related to the multiwinner voting problem within the field of social
choice~\citep{elkind2017properties,Brandt2016}. Similar to the
favorable properties of the Borda method~\citep{borda1781memoire,
  Brandt2016} in multiwinner voting, it also performs well in
portfolio selection when project costs are
uniform~\citep{ge2024knapsack}. One challenge in portfolio selection is
that agents may need to compare a potentially large number of
projects, especially when ranking them as is done in Borda
counting. Although this is an important problem in a variety of
applications, to date there has been little research on aggregation
methods based on pairwise comparisons, which provide a more practical
approach to ranking items.

Pairwise comparisons can help mitigate the cognitive burden associated
with directly ranking a large number of projects. The human short-term
memory is limited to processing around seven items at once (``Miller's
law'')~\citep{miller1956magical}, making ranking tasks inefficient as
the number of projects increases. Additionally, pairwise comparisons
have proven valuable in eliciting preferences and are essential in
cases where direct estimation of project value is difficult due to
psychological factors~\citep{torgerson1958theory}. In machine learning,
pairwise comparison is a well-studied problem, with research focusing
on reducing the complexity of comparisons needed to recover a full
ranking of $n$ items. For example, comparison methods that achieve an
expected lower bound of $\Omega(n)$ comparisons under certain
conditions have been proposed
\citep{pmlr-v28-wauthier13}. Additionally, active-learning techniques
that require no more than $\mathcal{O}(n\,\mathrm{polylog}(n))$
queries have been introduced \citep{ailon2012active}.

Most existing research on sorting assumes that comparison results are
consistent with the true underlying rankings. There has been
relatively little work on deriving rankings from noisy information,
such as in portfolio selection, where agents provide imprecise
estimates of project values. The authors of
\citep{DBLP:conf/soda/BravermanM08,braverman2009sorting} were among the
first to discuss algorithms for sorting based on noisy
information. These methods were later extended in
\citep{braverman2016parallel} to enable parallel processing. In this
paper, we propose pairwise-comparison rules based on the
Bradley--Terry model~\citep{zermelo1929berechnung, bradley1952rank},
which links rankings to win probabilities.

In Section~\ref{sec:related_work}, we give an overview of related
work, while in Section~\ref{sec:bradley_terry_model}, we present the
Bradley--Terry model and describe recent advances in its algorithmic
development. In Section~\ref{sec:portfolio_selection}, we connect the
Bradley--Terry model to portfolio-selection theory and describe
various methods, including a Quicksort approach, for
aggregating project evaluations from different agents in
Section~\ref{sec:aggregation_methods}. In
Section~\ref{sec:simulation_results}, we examine the performance of
these aggregation methods to show that several of the methods
  we propose perform better than existing aggregation methods. We
also demonstrate how our approach can be integrated with sampling
techniques to reduce the number of comparisons, alleviating cognitive
load on agents.  Finally, in Section~\ref{sec:conclusions}, we
summarize our findings and discuss our results.
\section{Related work}
\label{sec:related_work}
\paragraph{Bradley--Terry model}
The Bradley--Terry model is a statistical method used to rank $n$
items based on repeated pairwise comparisons. The model was introduced
in 1929 by Zermelo to study tournament outcomes
\citep{zermelo1929berechnung} and re-introduced in 1952 by the
eponymous Bradley and Terry~\citep{bradley1952rank}.  A common
application is ranking chess players according to their results in
matches against one another.  Due its versatility, the Bradley--Terry
model has also been applied to sports rankings, electoral preferences,
social choice modeling, psychological and healthcare studies, and to
other settings where relative comparisons are more practical than
isolated evaluations.  Many iterative algorithms have been proposed to
determine the maximum likelihood estimator parameters that can best
fit existing data \citep{Ford1957, Hunter2004} and to accelerate
algorithm convergence, such as Newman's iteration
\citep{newman2023efficient}.

Extensions include including ties between items, incorporating
ordering-based advantages (such as playing on one's home-field in
sports), or multiple (instead of pairwise) comparisons
\citep{Pendergrass1960, Rao1967, Davidson1970, Agresti1990,
  Marden1995}.  Bradley--Terry models are also used in machine
learning and reinforcement learning, as useful tools in ranking,
preference learning, learning from feedback, reward shaping, and other
problems involving human choices \citep{Muslimani2024, Wainer2023,
  Peschi2022}. They are also used to compare and rank large language
models (LLMs) through crowdsourced open platforms, or other expert
evaluators. Since direct pairwise LLM comparisons involve computations
that are $\mathcal{O}(n^2)$, novel comparison methods that use only a
subset of Bradley--Terry type pairings have been recently
introduced~\citep{Liusie2024}. Other modern applications include
comparing the ideological positions of US politicians using
specifically tuned LLMs~\citep{Wu2023} and ranking video-game players
based on skill so they can be properly paired in virtual matches. For
example, the proprietary TrueSkill ranking system utilizes a Bayesian
framework based on the Bradley--Terry model to incorporate
uncertainties in player skills. It can also be applied to matches with
more than two players \citep{Herbrich2006, Minka2018}.
\paragraph{Group decisions and portfolio selection}
A common problem in decision-theory is how to effectively aggregate
many individual inputs into a collective output. This issue arises in
voting systems, social choice problems, and organizational
decision-making.  Typical aggregation methods include treating all
inputs equally (\eg, by using the arithmetic mean), delegating to
specific individuals (at random or based on given criteria), using
majority rules, or biasing the outcome in favor of specific subgroups
\citep{einhorn1977quality}. Hierarchical team-decision making assumes
distributed expertise among tiered group members whose judgment is
aggregated using probabilistic methods \citep{Humphrey2002}. In
addition to developing computationally efficient aggregation methods,
research has also focused on other important aspects such as the
legitimacy of voting systems~\citep{hausladen2024voting} and their
capacity to mitigate polarization~\citep{alos2021voting}.

A decision problem that frequently arises in organizational settings
is that groups are tasked with selecting a subset of projects
from a portfolio. To model organizational decision-making,
agents can be considered as having specific expertise, while projects
are characterized by an intrinsic value (representing their long-term
benefit and assumed to be ground truth) and a defining
type~\citep{csaszar2013organizational}.  Agents do not know this
intrinsic value and must evaluate it; the larger the discrepancy
between project type and agent expertise, the larger the uncertainty
in the evaluation.  This model has been extended to include project
costs and budget limitations, and has been applied to social choice
problems \citep{boettcher2024selection, ge2024knapsack}.
\paragraph{Sorting methods}
Developing robust algorithms that operate effectively despite
unreliable or noisy information, without removing specific noisy
elements, is a problem that arises in many computing
applications~\citep{feige1994computing}. Unlike standard sorting
algorithms that focus on ordering a list of precisely known
values~\citep{Zopounidis2002}, comparing items with values affected by
various sources of uncertainty requires adapted sorting
approaches~\citep{DBLP:conf/soda/BravermanM08,braverman2009sorting,braverman2016parallel}. Several
strategies have been developed to improve on these so-called ``dirty''
comparisons, for example using the inaccurate results in parallel with
a subset of exact, ``clean'' comparisons to improve efficiency. The
number of the required clean comparisons can be tuned based on the
accuracy of the dirty comparisons \citep{DBLP:conf/nips/BaiC23,
  DBLP:conf/sosa/LuRSZ21, DBLP:conf/faw/ChanSW23}. In some
applications, it may be sufficient to use a relatively small number of
pairwise comparisons to obtain an approximate ranking. A lower bound
on the number of necessary pairwise comparisons has been derived
in~\citep{giesen2009}. Related work has also considered problems like
computing a longest increasing subsequence associated with a given
sequence of elements in the presence of comparison
errors~\citep{DBLP:conf/waoa/Geissmann18,geissmann2020longest}.
\section{The Bradley--Terry Model}
\label{sec:bradley_terry_model}
Here, we discuss the Bradley--Terry model in more mathematical detail.
The goal is to assign ``strength'' parameters to all players and rank
them accordingly.  The strength parameters determine the probability
of a win, tie, or loss when two players are placed in
competition. Specifically, if $\pi_i$ and $\pi_j$ represent the
strengths of competitors $i$ and $j$, respectively, the probability
that $i$ wins over $j$ is $\pi_i/(\pi_i + \pi_j)$.

Estimation of the strength parameters is usually performed using
maximum likelihood estimation. The basic idea is to maximize the
log-likelihood function of the observed competition outcomes to find
the most likely values of the players' strengths. Mathematically,
given a set of observed outcomes $w_{i j}$, where $w_{i j}$ is the
number of times competitor $i$ wins over competitor $j$, the
log-likelihood function for these outcomes under the Bradley--Terry
model is
\begin{equation}
  \begin{aligned}
  l(\boldsymbol{\pi}) = & \sum_{i \neq j} w_{i j}
  \ln \left( \frac{\pi_i}{\pi_i + \pi_j} \right) \\
  = & \sum_{i \neq j} w_{i j} \left[ \ln(\pi_i) - \ln(\pi_i + \pi_j) \right],
  \end{aligned}
    \label{eq:log_likelihood}
\end{equation}
where $\boldsymbol{\pi}=(\pi_1,\dots,\pi_n)^\top$ is the strength
parameter vector.

Maximizing the log-likelihood function $l(\boldsymbol{\pi})$ in
Eq.~\eqref{eq:log_likelihood} involves iterative updates of the
parameters $\boldsymbol{\pi}$. Zermelo
proved~\citep{zermelo1929berechnung} that this maximization has a
unique solution under certain conditions. The maximum can be found by
differentiating $l(\boldsymbol{\pi})$ with respect to
each parameter $\pi_i$ and setting the resulting expressions to zero,
resulting in an implicit expression for the strength parameters
\begin{equation}
    \pi_i = \frac{\sum_{j \neq i} w_{i j}}{\sum_{j \neq i} \frac{w_{i j} + w_{ji}}{\pi_i + \pi_j}}.
\label{maximum}
\end{equation}

Zermelo also proposed the first iterative
approach~\citep{zermelo1929berechnung} to solve Eq.~\eqref{maximum}. In
each iteration step, one calculates
\begin{equation}
    \pi_i' = \frac{\sum_{j \neq i} w_{i j}}{\sum_{j \neq i} \frac{w_{i j} + w_{ji}}{\pi_i + \pi_j}},
\end{equation}
where $\pi_i'$ denotes the updated strength of competitor $i$ starting
from strengths $\pi_i$ and $\pi_{j\neq i}$. This method is simple but
can be slow to converge. More recently, Newman proposed an alternative
iterative process \citep{newman2023efficient} that is substantially
faster than Zermelo's algorithm. His proposal is based on the
iteration
\begin{equation}
     \pi_i' = \frac{\sum_{j \neq i} \frac{w_{i j} \pi_j}{\pi_i + \pi_j}}{\sum_{j \neq i} \frac{w_{ji}}{\pi_i + \pi_j}},
\label{newman1}
\end{equation}
which converges faster than Zermelo's algorithm by a factor of $\sim
3--100$. Newman's iteration can be improved by incorporating updated
values after each iteration, enhancing both convergence speed and
stability as in the Gauss--Seidel method~\citep{kress2012numerical}. We
adapt this iteration process and represent it as
\begin{equation}
  \pi_i' = \frac{\sum_{j \neq i} \frac{w_{i j} \pi_j'}{\pi_i + \pi_j'}
    +\sum_{j > i} \frac{w_{i j} \pi_j}{\pi_i + \pi_j}}{\sum_{j \neq i} \frac{w_{ji}}{\pi_i + \pi_j'}+\sum_{j > i} \frac{w_{ji}}{\pi_i + \pi_j}}.
\label{newman2}
\end{equation}
Note that the strength parameters can become ill-defined if a player
never wins or never loses. For example, consider three players 1, 2,
and 3 with match results: 1 wins against 2, 1 wins against 3, and 2
wins against 3.  In such cases, the algorithm may converge to values
where $\pi_1$ grows towards infinity at a faster rate than $\pi_2$,
while $\pi_3$ converges to 0.
\begin{table}[h!]
\caption{An overview of the main model parameters. Unless otherwise
  stated, all parameters are real-valued.}  \centering
\begin{tabular}{ll}
\toprule
Symbol & Description \\ 
\midrule
$N\in\mathbb{Z}^+$ & Number of agents \\
$n \in\mathbb{Z}^+$ & Number of items (or projects) \\
$n^* \in\mathbb{Z}^{+}$ & Budget constraint\\ 
$i,j\in \{1, \dots, n\}$ & Project label \\
$\ell \in \{1, \dots, N \}$ & Agent label \\
$v_i \in \mathbb{R}^+$ & Value of project $i$ \\
$t_i\in[t_{\rm min}, t_{\rm max}]$ & Type of project $i$ \\
$e_\ell \in[e_{\rm min}, e_{\rm max}]$ & Expertise of agent $\ell$ \\
$\beta \geq 0 $ & Knowledge breadth of agents \\
$e_{\rm M}$ & Mean expertise level; \\ & $e_{\rm M} = (t_{\rm min} + t_{\rm max})/2$ \\
$v_{i\ell}$ & Value of project $i$ \\ & as evaluated by agent $\ell$ \\
$\eta_{i\ell} = v_{i\ell} - v_i$ & Noise in perceived value of project $i$ \\ & associated with agent $\ell$ \\
$\sigma_{i\ell} > 0 $ & Uncertainty of value of project $i$ \\ & associated with agent $\ell$ \\
$v'_i$ & Aggregate value of project $i$ \\ & over all $N$ agents \\
$w^{\ell}_{ij}\in(0,1)$ & Win probability of project $i$ \\ & outperforming project $j$ \\ & as predicted by agent $\ell$ \\
$W^\ell\in(0,1)^{n\times n}$ & Matrix of all win probabilities $w^{\ell}_{ij}$ \\ & as predicted by agent $\ell$ \\
$w'_{ij}\in(0,1)$ & Aggregated win probability of \\ & project $i$ outperforming project $j$ \\
$W'\in(0,1)^{n\times n}$ & Matrix of all aggregated \\ 
& win probabilities $w'_{ij}$ \\
\bottomrule
\end{tabular}
\label{tab:overview}
\end{table}
\section{Portfolio Selection}
\label{sec:portfolio_selection}
To model the selection of projects from a portfolio under cost
constraints, we build on the framework proposed in
\citep{boettcher2024selection}. Each project $i\in\{1,\dots,n\}$ is
characterized by two parameters: its type $t_i\in[t_{\rm min}, t_{\rm
    max}]$ and value $v_i\in\mathbb{R}^+$.  In the project selection
context, the values $v_i$ define the true benefit of project $i$ over
a specific time horizon, if chosen. The true benefit may evolve and be
uncertain over time due to societal value shifts, environmental
changes, and complex interactions with other selected projects $j \neq
i$. We do not consider these sources of uncertainty in $v_i$ (which is
here considered the ground truth) and restrict ourselves to each
agent's uncertainty in the estimation of $v_i$ at the time of
evaluation. This leads to subjective evaluations $v_{i \ell}$ of
project $i$ from each agent $\ell\in\{1,\dots,N\}$. To represent $v_{i
  \ell}$ we first assume that each agent $\ell$ involved in the
decision-making process has a level of expertise $e_\ell\in[e_{\rm
    min}, e_{\rm max}]$ given by
\begin{equation}
	e_\ell = e_{\rm M} - \frac{N + 1 - 2 \ell}{N-1} \beta.
\label{expert}
\end{equation}
According to Eq.~\eqref{expert} the $e_\ell$ values are evenly spaced
across the interval $[e_{\rm M} - \beta, e_{\rm M} + \beta]\equiv
[e_{\rm min}, e_{\rm max}]$. Here, $e_{\rm M}$ represents the mean
expertise level and $\beta$ denotes the knowledge breadth that
determines the expertise spread.  For mathematical convenience, we set
$e_{\rm M}=(t_{\rm min}+t_{\rm max})/2$ so that the mean expertise
coincides with the mean project type.

The values $t_i$ and $e_\ell$ do not have any specific meaning; they
are simply labels used to differentiate between various types and
expertise levels. However, the alignment between $t_i$ and $e_\ell$
affects the accuracy of $v_{i \ell}$, agent $\ell$'s evaluation of
project $i$'s value.\footnote{Our approach to describing
domain-specific expertise is similar to that used in Hotelling models,
where preferences are represented as distances along a
line~\citep{hotelling1929stability,NOVSHEK1982199}.} Specifically, we
assume that the noise in the ``perceived value,'' $\eta_{i\ell} =
v_{i\ell} - v_i$, follows a normal distribution with standard
deviation $\sigma_{i\ell} = |t_i - e_\ell|$.  That is, $\eta_{i\ell}
\sim \mathcal{N}(0,\sigma_{i\ell}^2)$, meaning that the closer the
agent's expertise is to the project type, the lower the
uncertainty. Each project is evaluated by $N$ agents, and their
aggregated preferences determine the final selection. Given a limited
amount of resources, only a fixed number $n^* < n$ projects can be
selected.

In this paper, we extend the described model of portfolio selection to
pairwise comparisons between projects. Suppose agent $\ell$ is
evaluating projects $i$ and $j$, with perceived values $v_{i\ell}$ and
$v_{j\ell}$, respectively, and corresponding noise terms
$\eta_{i\ell}$ and $\eta_{j\ell}$. How would this agent evaluate the
win probability $w_{ij}^\ell$ that project $i$ is better than project
$j$? As a starting point, we define
\begin{align}
\begin{split}
    w_{ij}^\ell &\coloneqq \Pr\left(v_i > v_j \right) = \Pr\left((v_{i\ell}-\eta_{i\ell})>(v_{j\ell}-\eta_{j\ell})\right) \\
    &=\Pr\left((\eta_{i\ell}-\eta_{j\ell})<(v_{i\ell}-v_{j\ell})\right).
\end{split}
\label{eq:wijl}
\end{align}
Under the assumption that the noise in the perceived value is
independently and normally distributed, the quantity $\eta_{i\ell} -
\eta_{j\ell}$ follows a normal distribution with a mean of 0 and a
standard deviation of $\sqrt{\sigma_{i\ell}^2 + \sigma_{j\ell}^2}$. We
can thus rewrite Eq.~\eqref{eq:wijl} as
\begin{equation}
    w^{\ell}_{ij} = \Phi \left( \frac{v_{i\ell} - v_{j\ell}}{\sqrt{\sigma_{i\ell}^2 + \sigma_{j\ell}^2}} \right),
\label{pairwise_comparison}
\end{equation}
where $\Phi$ is the cumulative distribution function of the standard
normal distribution. An immediate consequence of the above equation is
that $w^{\ell}_{ij} = 1 - w^{\ell}_{ji}$.

In Table~\ref{tab:overview}, we provide an overview of the main model
parameters used in this work. Some parameters, such as aggregated
values and win probabilities, will be introduced in the next section,
where we discuss various aggregation methods for identifying the most
valuable projects within a given portfolio based on their performance
$E (\beta; N, n, n^*)$. This quantity is defined as the expected value
over $n^*$ out of $n$ projects that are evaluated by $N$ agents with
knowledge breadth $\beta$. We compute the expected value over a given
type distribution.

As an example, we consider $N=3$ agents, each with a knowledge
  breadth of $\beta=0$, and $n=3$ projects, from which $n^*=2$ must be
  selected. The project values are $v_1=1$, $v_2=2$, and $v_3=3$. We
  assume that agents perceive the true project values (\ie,
  $v_{i\ell}=v_i$ for $\ell\in\{1,2,3\}$). The performance for this
  example is calculated as $E(\beta=0; N=3, n=3, n^*=2)=v_2+v_3=5$.

Some of the aggregation approaches that we study in this work
will be based on project value estimates $v_{i\ell}$ while others will
employ win probabilities $w^{\ell}_{ij}$. We will show that
aggregation methods using win probabilities instead of value estimates
typically perform better.
\section{Aggregation Methods}
\label{sec:aggregation_methods}
In the portfolio-selection model that we consider in this work,
projects are chosen based on information collected from multiple
agents. This information contains noise, and a key challenge is to
design aggregation methods that effectively integrate all the agents'
inputs to maximize the expected value of the selected projects. One
possibility is to aggregate the estimated values provided by the
agents using the arithmetic mean. However, value estimates may be
difficult to ascertain in practice. Additionally, outliers can easily
bias the arithmetic mean towards inaccurate value estimates. Previous
research~\citep{boettcher2024selection} has shown that a ranking-based
method using Borda scores is more robust to outliers than the
value-based arithmetic mean. Here, we leverage the win
  probabilities expressed in Eq.~\eqref{pairwise_comparison} and
  incorporate them into existing aggregation methods. The main advantage
of this method is that rankings can be predicted directly from win
probabilities that are associated with pairwise comparisons of
projects, so that project selection can proceed without relying on
explicit value estimates.

We will proceed with an overview of the aggregation methods that we
employ in this work.
\subsection{Overview}
\paragraph*{a) Arithmetic Mean} This method requires all agents to provide precise perceived values $v_{i\ell}$, which are then averaged using the arithmetic mean to obtain the aggregated value $v_{i}'$. That is,
\begin{equation}
    v'_i = \frac{1}{N} \sum_{\ell=1}^{N}v_{i\ell}. 
    \label{eq:arithmetic_mean}
\end{equation}
The $n^*$ projects with largest aggregate values $v'_i$ are then selected.
\paragraph*{b) Borda Count} The Borda Count is based on the eponymous method introduced by Jean-Charles de Borda in the late 18th century \citep{borda1781memoire}. In this approach, each agent $\ell$ ranks the $n$ projects in descending order based on their perceived values $v_{i\ell}$. For each project $i$, we denote its position in agent $\ell$'s preference list by $\mathrm{pos}_\ell(i)$. The aggregated score $s_i$ for project $i$ is then calculated as the sum of the reversed ranks across all $N$ agents. That is,
\begin{equation}
    s_i = \sum_{\ell=1}^{N} (n - \text{pos}_\ell(i)).
    \label{eq:Borda}
\end{equation}
The top $n^*$ projects with the highest aggregated scores are
selected. According to \citep{boettcher2024selection}, this method is
particularly robust against misclassification and often outperforms
the Arithmetic Mean, especially in scenarios with high uncertainty.
\paragraph*{c) Quicksort} Quicksort is a widely used sorting algorithm,
first introduced in \citep{hoare1962quicksort}, which employs a
divide-and-conquer approach to sort elements. Its average-case time
complexity is $\mathcal{O}(n\, \log(n))$, making it one of the most
efficient sorting algorithms to date
\citep{cormen2022introduction}. When applied to project selection, as
demonstrated in Algorithm \ref{quick_sort_alg}, the algorithm selects
a ``pivot'' project from the middle of the list of available projects
and partitions the remaining projects into two sub-lists: one
containing projects ranked worse than the pivot, and the other
containing projects ranked better than or equal to the pivot. This
partitioning process is recursively applied to each sub-list. In our
implementation, a project is considered better than the pivot if its
aggregated win probability against the pivot exceeds 0.5. The
aggregated win probability associated with projects $i$ and $j$ is
\begin{equation}
    w'_{ij} = \frac{1}{N}\sum_{\ell=1}^N w^{\ell}_{ij},
\label{eq:wij_mean}
\end{equation}
where $w^{\ell}_{ij}$ is given by Eq.~\eqref{pairwise_comparison}.

The Quicksort method produces a list of projects ranked based on their
aggregated win probabilities in ascending order, from which the last
$n^*$ projects are selected.

\begin{algorithm}[t]
\caption{Quicksort with aggregated win-probability matrix}
\begin{algorithmic}[1]
\REQUIRE Aggregated win-probability matrix $W'$ of size $n\times n$
\ENSURE Sorted index array $idx$
\STATE $idx \leftarrow$ list of integers from $0$ to $n-1$

\STATE \textbf{function} \textsc{Partition}($low$, $high$)
    \STATE $i \leftarrow low - 1$
    \FOR{$j \leftarrow low$ \TO $high - 1$}
        \IF{$W'[idx[j], idx[high]] < 0.5$}
            \STATE $i \leftarrow i + 1$
            \STATE Swap($idx[i]$, $idx[j]$)
        \ENDIF
    \ENDFOR
    \STATE Swap($idx[i + 1]$, $idx[high]$)
    \STATE \textbf{return} $i + 1$
\STATE \textbf{end function}

\STATE \textbf{function} \textsc{QuickSortRecursive}($low$, $high$)
    \IF{$low < high$}
        \STATE $pi \leftarrow$ \textsc{Partition}($low$, $high$)
        \STATE \textsc{QuickSortRecursive}($low$, $pi - 1$)
        \STATE \textsc{QuickSortRecursive}($pi + 1$, $high$)
    \ENDIF
\STATE \textbf{end function}

\STATE \textsc{QuickSortRecursive}($0$, $n - 1$)
\STATE \textbf{return} $idx$
\end{algorithmic}
\label{quick_sort_alg}
\end{algorithm}
\paragraph*{d) Bradley--Terry Method} The Bradley--Terry model is
usually employed in tournament settings, where the quantities $w_{ij}$
are integers and represent the number of times that competitor $i$
wins over competitor $j$. We use the Bradley--Terry model to devise a
portfolio selection method that includes real-valued probabilities
$w_{ij}^\ell$ as follows:
\begin{itemize}
    \item First, each agent $\ell$ provides their predicted
      probabilities $w_{ij}^\ell$ for all pairwise comparison
      results. We use $W^{\ell}\in(0,1)^{n\times n}$ to denote the
      corresponding win-probability matrix.
    \item Then, aggregated win probabilities $w'_{ij}$ are computed
      according to Eq.~\eqref{eq:wij_mean}. We use
      $W'\in(0,1)^{n\times n}$ to denote the corresponding aggregated
      win-probability matrix.
    \item Next, Newman's iteration is used to determine the relative
      strength of each project using Eqs.~\eqref{newman1} and
      \eqref{newman2}.
    \item Finally, projects are selected in descending order of
      relative strength until the desired number of projects $n^*$ is
      reached.
\end{itemize}
Since it may not be feasible for each agent to perform pairwise
comparisons for all projects in the first step, we propose sampling
approaches in Section~\ref{samplingschemes} so that only a subset of
pairwise comparisons are performed. In the second step, one may
consider win-probability aggregation methods different from
Eq.~\eqref{eq:wij_mean}.

Given that the aggregation methods presented here involve different
quantities (\ie, values, scores, and win probabilities), we will now
discuss some advantages and pitfalls associated with the (a-d) methods
outlined above.
\subsection{Values, scores, or win probabilities?}
\label{sec:example}
Aggregating win probabilities using Eq.~\eqref{eq:wij_mean} offers an
advantage over employing the Arithmetic Mean as per
Eq.~\eqref{eq:arithmetic_mean}, particularly when handling outliers in
project-value evaluations. To illustrate this point, we consider three
agents evaluating two projects, Project 1 and Project 2. The first
agent holds a highly favorable view of Project 1, while the other two
agents assign lower value estimates to it. If the first agent's
evaluation is an outlier\textemdash say $v_{11}$ approaches
infinity\textemdash this outlier's effect differs significantly
between the two methods.

With the arithmetic mean, the aggregated value for Project 1, $v'_1$,
becomes highly skewed by the outlier and may approach infinity as
well. This disproportionate influence from a single agent distorts the
collective assessment of Project 1's value.

In contrast, when using the win probability aggregation method, the
outlier's impact is mitigated. We assume that the extreme value from
the first agent translates into a win probability of $w_{12}^1 =
0.98$, indicating a strong preference. If the other two agents provide
negative assessments of Project 1 with respect to Project 2, such as
$w_{12}^2 = w_{12}^3 = 0.2$, the aggregated win probability,
calculated using Eq.~\eqref{eq:wij_mean}, results in $w_{12}' =
0.46$. This result is more closely aligned with the agents'
evaluations than the one obtained with the Arithmetic Mean shown in
Eq.\,\eqref{eq:arithmetic_mean}.

Using win probabilities also offers an advantage over the Borda Count,
as it more precisely captures individual preferences through
real-valued probabilities. For example, consider two agents evaluating
two projects, Project 1 and Project 2. The first agent strongly
prefers Project 1 over Project 2 ($w_{12}^1 = 0.8$), while the second
agent only slightly favors Project 2 over Project 1 ($w_{12}^2 =
0.46$). The aggregated win probability, $w_{12}' = 0.63$, indicates
that Project 1 is the preferred choice overall, reflecting the
stronger preference of the first agent. This approach takes into
account the intensity of each agent's preference.

On the other hand, if the Borda Count is used, each project would
receive a Borda score of 1, resulting in a tie. This outcome fails to
differentiate between the strong preference expressed by the first
agent and the more moderate preference of the second agent.
\subsection{Sampling pairwise comparisons}
\label{samplingschemes}
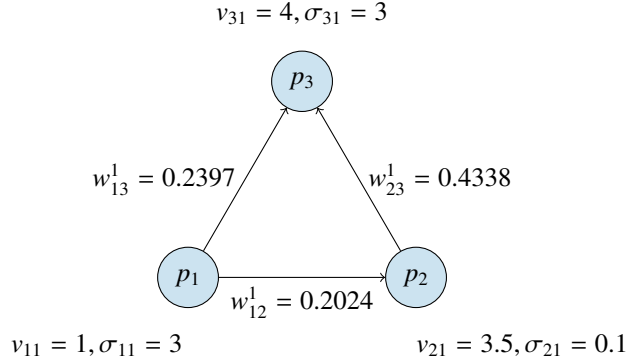
\begin{figure}[t]
    \centering
    \input{network.tikz}
    \caption{Comparison of three projects $p_1$, $p_2$, and $p_3$ by a
      single agent. Each node represents a project and each directed
      edge represents a pairwise comparison between projects.}
    \label{fig:network}
\end{figure}
When applying the Bradley--Terry method to portfolio selection,
sampling pairwise comparisons (\ie, selecting a subset of win
probabilities $w_{ij}^\ell$) can be a cost-effective strategy in
practical implementations, as each additional comparison requires more
resources. Moreover, we will show that incorporating sampling
protocols into aggregation methods can improve performance.

To illustrate how different methods of comparing projects can
influence the resulting rankings, consider a simple example involving
a single agent evaluating three projects. The value estimates for
these projects are $v_{11} = 1$, $v_{21} = 3.5$, and $v_{31} = 4$,
with corresponding uncertainties $\sigma_{11} = 3$, $\sigma_{21} =
0.1$, and $\sigma_{31} = 3$. Depending on the sampling strategy
employed, these pairwise comparisons can yield different rankings (see
Figure~\ref{fig:network}). For instance, comparing Project 1 with
Project 2 results in a win probability of $w^1_{12} = 0.2024$, while
comparing Project 2 with Project 3 yields $w^1_{23} = 0.4338$. This
sequence of comparisons leads to the ranking: Project 3 $\succ$
Project 2 $\succ$ Project 1, where $x\succ y$ indicates that $x$ is
strictly preferred over $y$. However, if we compare Project 1 with
both Project 2 and Project 3, the win probabilities $w^1_{12} =
0.2024$ and $w^1_{13} = 0.2397$ produce a different ranking: Project 2
$\succ$ Project 3 $\succ$ Project 1. Thus, different methods of
performing pairwise comparisons can lead to varying rankings.

Instead of performing all $\mathcal{O}(n^2)$ comparisons, we employ an
$\mathcal{O}(n)$ cyclic graph sampling method. This technique can be
visualized as extracting a subgraph from the complete graph generated
by $n$ projects. Given an ordered list of all $p_i$ projects such as
$(p_1, p_2, \ldots, p_{n})$, the cyclic graph sampling of pairwise
comparisons is defined as
\begin{equation}
    ((p_1, p_2), (p_2, p_3), \ldots, (p_{n-1}, p_{n}), (p_{n}, p_1)).   
\label{eq:cyclic_graph_sampling}
\end{equation}
In this notation, the pairs in parentheses represent the pairwise
comparisons conducted between these projects.

The performance of the cyclic graph sampling approach is influenced by
how projects are initially ordered in the list. In this work, we
employ a two-phase approach. In the first phase, an initial
project-ranking approximation can be obtained through random sampling
or by using existing ranking algorithms. This preliminary ranking then
serves as the input for the second phase, where cyclic graph sampling
is employed to refine the rankings through an optimization step based
on the Bradley--Terry method.

We propose two additional aggregation methods using the cyclic graph
sampling shown in Eq.~\eqref{eq:cyclic_graph_sampling}. Both are
based on a two-phase approach and are discussed below
\paragraph*{e) Two-Phase Bradley--Terry method}
\begin{itemize}
    \item In the first phase, we begin by generating an initial
      ranking using a list in which each project $p_i$ ($i \in {1,
        \dots, n}$) is selected uniformly at random without
      replacement from the $n$ available projects. Next, the values of
      $w'_{ij}$ are calculated via Eq.~\eqref{eq:wij_mean} according
      to the cyclic graph sampling of the randomly ordered list as
      shown in \eqref{eq:cyclic_graph_sampling}. Finally, Newman's
      iteration illustrated in Eqs.~\eqref{newman1} and
      \eqref{newman2} is applied to the win probabilities $w'_{ij}$ to
      obtain an approximate ranking.
    \item In the second phase, starting from the approximate ranking,
      we compute the corresponding win probabilities $w'_{ij}$ using
      cyclic graph sampling \eqref{eq:cyclic_graph_sampling}. To
      further refine the ranking, we apply Newman's iteration again,
      while also incorporating the win probabilities obtained in the
      first phase. Win probabilities that were not calculated in
      either phase one or phase two are set to 0.
\end{itemize}
\paragraph*{f) Two-Phase Quicksort}
\begin{itemize}
    \item In the first phase, we approximate the project ranking using
      the Quicksort algorithm. Instead of relying on randomly selected
      pairwise comparisons, here we apply the Quicksort algorithm to
      the matrix of aggregated win probabilities $W'$, with elements
      $w'_{ij}$ (as shown in Eq.~\eqref{eq:wij_mean}), to generate an
      initial ranking of items. During this step, only the necessary
      entries of the aggregated win probabilities $W'$ are sampled,
      resulting in an $\mathcal{O}(n\,\log(n))$ complexity. In
      addition to calculating $w'_{ij}$, the Quicksort algorithm also
      produces an initial ranking. However, because the underlying
      perceived values are noisy observations, the output of the
      Quicksort algorithm does not represent the true ranking as it
      would in the absence of uncertainty.
    \item In the second phase, starting from the Quicksort ranking, we
      compute the corresponding win probabilities $w'_{ij}$ using
      cyclic graph sampling \eqref{eq:cyclic_graph_sampling}. Newman's
      iteration shown in Eqs.~\eqref{newman1} and \eqref{newman2}) is
      then applied to determine a refined ranking. Unlike in the
      Two-Phase Bradley--Terry method, we only consider win
      probabilities associated with the cyclic graph structure and not
      those obtained in the first phase.
\end{itemize}
\section{Simulation Results}
\label{sec:simulation_results}
We now compare the effectiveness of the aggregation methods (a--f) in
achieving a high expected value for the selected projects, as
quantified by the performance measure $E(\beta; N, n, n^*)$. This is a
measure that quantifies the expected value of the $n^*$ (out of $n$)
selected projects, each evaluated by $N$ agents with with knowledge
breadth $\beta$. Recall that the knowledge breadth determines the
spread in agents' expertise according to Eq.~\eqref{expert}. Following
prior
work~\citep{csaszar2013organizational,boettcher2024selection,ge2024knapsack},
we calculate the expected value over the project-type distribution
$\mathcal{U}(0,10)$. The expertise value of the central decision maker
is set at $e_{\rm M}=(t_{\rm min}+t_{\rm max})/2=5$. In our
simulations, we consider a scenario with $n=30$ projects, $N=3$
agents, and a target of selecting $n^*=15$ projects. We define the
value of project $i$ as $v_i = i$ ($i\in\{1,\dots,30\}$). The
uncertainty in agent $\ell$'s project evaluations is quantified by
additive Gaussian noise with zero mean and standard deviation
$\sigma_{i\ell}=|t_i - e_{\ell}|$, where the expertise level
$e_{\ell}$ of agent $\ell$ is given by
Eq.~\eqref{expert}. Prior work~\citep{boettcher2024selection}
  has demonstrated that variations in value distribution, type
  distribution, and other parameters rarely affect the relative
  ordering of aggregation-rule performance.

All our results are based on Monte Carlo simulations. For methods
based on pairwise comparisons and win probabilities, we use 100,000
independent and identically distributed samples. For the remaining two methods, Arithmetic Mean and Borda Count, which are computationally less demanding, we increase the sample size to 500,000. The
theoretical maximum performance is $\sum_{i=16}^{30} v_i =
\sum_{i=16}^{30} i = 345$.

We consider two scenarios for computing the win probabilities
$w'_{ij}$. In the first scenario, the probabilities are calculated
according to Eqs.~\eqref{pairwise_comparison} and
\eqref{eq:wij_mean}. However, in real-world applications of
aggregation methods based on pairwise comparisons and win
probabilities, assigning probabilities with several decimal places may
be impractical. Therefore, in the second scenario, we prespecify a set
of win probabilities from which agents can choose when making pairwise
comparisons.
\subsection{Continuous win probabilities}
\begin{figure}[t]
    \centering
    \includegraphics[width=0.75\linewidth]{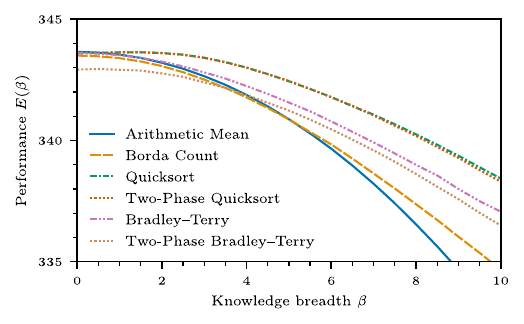}
    \caption{Portfolio selection with $n=30$ projects, $N=3$ agents,
      and a target of selecting $n^*=15$ projects. We show the
      performance $E(\beta; N, n, n^*)$ as a function of the knowledge
      breadth $\beta$ for the six aggregation methods (a--f).}
    \label{fig:bradley_terry}
\end{figure}
In Figure~\ref{fig:bradley_terry}, we show the performance $E(\beta;
N=3, n=30, n^*=15)$ for the six aggregation methods (a--f) as a
function of knowledge breadth $\beta$. Prior
work~\citep{boettcher2024selection} has highlighted that both the
Arithmetic Mean and the Borda Count are effective in identifying
high-value projects within a portfolio. In particular, the Borda Count
is more robust to evaluation outliers than the Arithmetic Mean and
performs well across a wide range of model parameters. Our results in
Figure~\ref{fig:bradley_terry} show that Quicksort (c), Two-Phase
Quicksort (f), and the Bradley--Terry method using all pairwise
comparisons (d), outperform both the Arithmetic Mean (a) and Borda
Count (b) methods, particularly at higher values of $\beta$. The
Two-Phase Bradley--Terry method (e), which employs a cyclic graph
sampling approach of pairwise comparisons, performs worse than both
the Arithmetic Mean (a) and Borda Count (b) for knowledge breadths
$\beta \lesssim 5.5$.

As a robustness check, we also conducted simulations for $N=15$ and
$N=30$ agents. We observed that the performance of all methods and
observed that for all six methods the performance increases with
increasing $N$, while their relative performance remains
similar. Additionally, the performance gap between the Two-Phase
Bradley--Terry method and the other methods widens. This is because
the Two-Phase Bradley--Terry method uses a sampling protocol that
leaves more entries in the aggregated win probability matrix $W'$
empty, compared to other methods based on win probabilities.
\subsection{Discrete win probabilities}
\begin{figure}[t]
    \centering
    \includegraphics[width=0.75\linewidth]{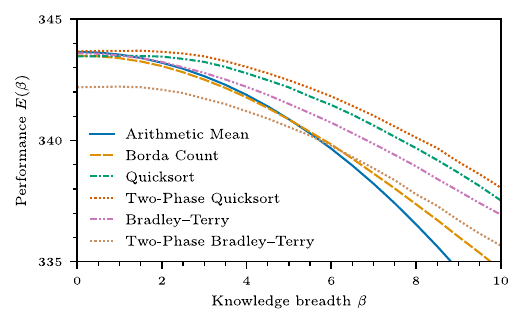}
    \caption{Portfolio selection with $n=30$ projects, $N=3$ agents,
      and a target of selecting $n^*=15$ projects. We show the
      performance $E(\beta; N, n, n^*)$ as a function of the knowledge
      breadth $\beta$ for the six aggregation methods (a--f). In all
      approaches that are based on win probabilities, agents select
      one of the following values $\{0.01, 0.1, 0.2, \ldots, 0.9,
      0.99\}$.}
    \label{fig:bradley_terry_discrete probability}
\end{figure}
In practical applications of the Bradley--Terry method, it may be
necessary to prespecify a set of win probabilities $w^{\ell}_{ij}$
from which agents can choose. For example, one could use a set of
prespecified probabilities such as $\{0.01, 0.1, 0.2, \ldots, 0.9,
0.99\}$. This approach simplifies the win-probability values by
limiting them to a finite and manageable set, which is useful in
decision-making scenarios where achieving a high degree of precision
is not feasible.

In Figure~\ref{fig:bradley_terry_discrete probability}, we show a
comparison of the aggregation methods (a--f) where the win
probabilities are restricted to values taken from the set $\{0.01,
0.1, 0.2, \ldots, 0.9, 0.99\}$. The relative performance ranking of
the methods remains unchanged. However, the performance values of
Quicksort and Two-Phase Quicksort exhibit a greater difference
compared to the continuous case shown in
Figure~\ref{fig:bradley_terry}. Recall that the Two-Phase Quicksort
method employs a refinement phase in which the final ranking is
computed according to Newman's iteration (see Eqs.~\eqref{newman1} and
\eqref{newman2}). While this second phase had little impact on
performance in the continuous case, it substantially affected results
when using the prespecified win probabilities listed above. This
aligns with the intuition that Newman's iteration (or similar
iterative methods used in the Bradley--Terry method) performs well in
scenarios where rankings are derived from a limited set of tournament
outcomes.

The Arithmetic Mean and Borda Count methods rely on value estimates
and scores, respectively, rather than win probabilities. Assigning
value estimates can, in principle, be done directly without comparing
projects. However, assigning ranking scores, as in the Borda Count,
requires project comparisons. In our implementation of the Borda
Count, we sort the agents' value estimate lists to generate the final
rankings based on Eq.~\eqref{eq:Borda}. When using Quicksort to obtain
the scores, the average number of comparisons is
$\mathcal{O}(n\,\log(n))$.

\begin{figure}[t]
    \centering
    \includegraphics[width=0.75\linewidth]{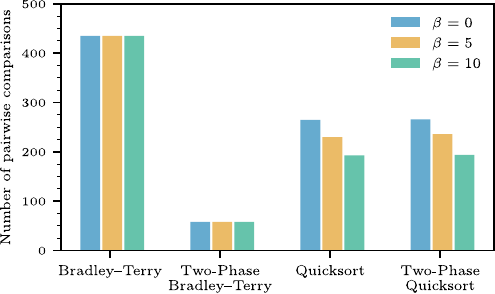}
    \caption{Number of pairwise project comparisons across aggregation
      methods (c--f) for knowledge breadths $\beta \in \{0, 5, 10\}$
      used to determine the performance $E(\beta)$ in
      Figure~\ref{fig:bradley_terry_discrete probability}.}
    \label{fig:bar_plot}
\end{figure}

For the remaining methods that are based on win probabilities, the
average number of comparisons are as follows:
\begin{itemize}
    \item Bradley--Terry (all pairwise comparisons): $\mathcal{O}(n^2)$
    \item Two-Phase Bradley--Terry: $\mathcal{O}(n)$
    \item Quicksort (without a second refinement phase): $\mathcal{O}(n\,\log(n))$
    \item Two-Phase Quicksort: $\mathcal{O}(n\,\log(n))$
\end{itemize}
In Figure~\ref{fig:bar_plot}, we show the number of pairwise
comparisons for each of these four approaches. The standard
Bradley--Terry aggregation method considers all $30(30-1)/2 = 435$
possible comparisons, regardless of the value of $\beta$. The
Two-Phase Bradley--Terry method involves approximately 58 comparisons
for the given values of $\beta$. For Quicksort without a second
refinement phase, the number of pairwise comparisons decreases from
265 for $\beta = 0$ to 193 for $\beta = 10$. The Two-Phase Quicksort
approach results in slightly more comparisons, with 266 for $\beta =
0$ and 194 for $\beta = 10$.

The observed decrease in the number of pairwise comparisons results
from the interplay between noise in the perceived values and the
divide-and-conquer nature of Quicksort. The algorithm compares
elements against a pivot to split the dataset into two parts. The
fastest case occurs when the two parts contain an equal number of
projects, while the slowest case happens when one part contains all
the remaining projects, and the other part contains none. As the noise
in project values increases with $\beta$, it becomes less likely that
highly imbalanced sublists arise during Quicksort
recursion. Therefore, the algorithm is expected to be more efficient
for large values of $\beta$.

Although the Bradley–Terry, Quicksort, and Two-Phase Quicksort
  methods demonstrate the best performance in the simulations
  considered, the number of pairwise comparisons they require is
  likely too high for practical applications. In contrast, the
  Two-Phase Bradley–Terry method achieves favorable performance with a
  relatively small number of pairwise comparisons, making it the most
  practical approach for real-world use.
\section{Conclusions}
\label{sec:conclusions}
In this work, we compared and contrasted six aggregation methods (a-f)
for portfolio selection of projects with uncertain values. Of these,
four novel methods (c-f) are based on pairwise comparisons. Agents are
tasked with selecting a subset of available projects. The accuracy of
their evaluations improves when the agents' expertise aligns well with
the project types. However, when there is a misalignment between
expertise and project types, evaluations are more prone to errors.

Agents may assign estimated values to projects, which can then be
aggregated to make final decisions about which projects to
select. When value estimates are difficult to ascertain, or when there
is a risk of outlier evaluations due to mismatches between expertise
and project types, it may be more appropriate to use Borda-type
methods. These methods rank projects based on their perceived value,
and the rankings are aggregated. However, ranking projects can also be
challenging, especially when agents must evaluate a large number of
projects relative to each other.

To address this challenge, we established a connection between
portfolio selection and the Bradley--Terry model in which rankings of
items are derived from pairwise comparisons and the associated win
probabilities. We proposed two main aggregation methods based on win
probabilities. The first method uses an extension of Quicksort to
produce project rankings based on aggregate win probabilities, with a
computational complexity of $\mathcal{O}(n \log n)$. The second
approach employs Newman's iteration to compute rankings from a set of
pairwise comparisons and their corresponding win probabilities. To
reduce the number of comparisons, we introduced a cyclic graph
sampling method, which achieved favorable performance with
$\mathcal{O}(n)$ comparisons instead of the $\mathcal{O}(n^2)$
required for all possible pairwise comparisons. Similar
  graph structures have also been studied in the context of
  subgraph matching~\citep{ge2025iterative}.

The methods we propose have applications in participatory budgeting,
social choice, organizational decision-making, and other resource
allocation problems involving decision-making under
uncertainty. Furthermore, our sampling and ranking methods can
  be effectively applied to benchmark foundation models such as
  LLMs~\citep{zhang2024inherent,ghosh2024onebench,Liusie2024}.

An interesting direction for future research is to extend the proposed
Bradley--Terry aggregation methods by incorporating delegation
strategies, where only agents with suitable expertise are queried, or
by using approaches based on the median instead of the arithmetic
mean. A related promising direction is to examine how
  interactions between agents shape their evaluations, using models of
  social influence on networks~\citep{stomakhin2011reconstruction,bottcher2017critical,zipkin2016point}.

Another avenue for future work is the study of sampling methods akin
to our cyclic graph sampling method that can achieve good performance
with a subset of all pairwise comparisons. Identifying ways to
sparsify the aggregate win probability matrix $W'$ can make our
proposed method more practical.

Finally, our work focused on uniform project costs, a single type
distribution, and a single set of project values. These assumptions
could be relaxed to examine the impact of heterogeneous project costs
and of other type and value distributions. A broader analysis along
these lines would provide further insight into the performance of
aggregation methods based on pairwise comparisons and corresponding
win probabilities.

\section*{Acknowledgments}
The authors thank Ronald Klingebiel for helpful discussions. This work
was supported by the Army Research Office through W911NF-23-1-0129
(YG, LB, MRD) and by the National Science Foundation through grant
MRI-2320846 (MRD). LB also acknowledges funding from hessian.AI.

\bibliographystyle{elsarticle-num}
\bibliography{main}

\end{document}

%% file: network.tikz
\begin{tikzpicture}

\node[circle, draw, fill=bluecolorblind!20, inner sep=2pt, minimum size=0.8cm] (1) at (0, 0) {$p_1$};
\node[circle, draw, fill=bluecolorblind!20, inner sep=2pt, minimum size=0.8cm] (2) at (3, 0) {$p_2$};
\node[circle, draw, fill=bluecolorblind!20, inner sep=2pt, minimum size=0.8cm] (3) at (1.5, 2.6) {$p_3$};

\draw[->] (1) -- (2) node[midway, below] {$w^1_{12} = 0.2024$};
\draw[->] (2) -- (3) node[midway, right] {$w^1_{23} = 0.4338$};
\draw[->] (1) -- (3) node[midway, left] {$w^1_{13} = 0.2397$};

\node at (-1.2, -0.9) {$v_{11} = 1, \sigma_{11} = 3$};
\node at (4.4, -0.9) {$v_{21} = 3.5, \sigma_{21} = 0.1$};
\node at (1.5, 3.5) {$v_{31} = 4, \sigma_{31} = 3$};

\end{tikzpicture}